%% file: cair2017-ws-julia.tex
\documentclass[sigconf,natbib]{acmart}

\usepackage{booktabs} 
\usepackage{xspace}
\usepackage{longtable}
\usepackage{amssymb,amsmath}
\usepackage{bm}
\usepackage{amsfonts}
\usepackage{multirow}
\usepackage[normalem]{ulem}
\usepackage{wrapfig}
\usepackage{paralist}

\newcommand{\is}{interaction signals\xspace}

\newcommand{\ias}{personal assistant\xspace}
\newcommand{\iass}{personal assistants\xspace}
\newcommand{\mi}{mobile interaction\xspace}
\newcommand{\device}{gesture- and voice-controlled devices\xspace}
\newcommand{\vgbs}{voice- and gesture-based signals\xspace}
\newcommand{\gbs}{gesture-based signals\xspace}

\newcommand{\gesture}{gesture-based\xspace}
\newcommand{\voice}{voice-based\xspace}
\newcommand{\sat}{user satisfaction\xspace}
\newcommand{\casat}{context-aware user satisfaction\xspace}

\newcommand{\rqmain}{\textbf{Our aim} is to exploit \textsl{\vgbs}, trackable at large scale, for understanding \textsl{\casat} with \iass.\xspace}
\newcommand{\rqone}{\textbf{RQ1}: \textsl{How to model interaction with \device?}\xspace}
\newcommand{\rqtwo}{\textbf{RQ2}: \textsl{How to define \casat with \iass in mobile environments?}\xspace}
\newcommand{\rqthree}{\textbf{RQ3}: \textsl{How to predict \casat with \iass using \gbs on mobile devices?}\xspace}

\setcopyright{rightsretained}

\acmDOI{10.475/123_4}

\acmISBN{123-4567-24-567/08/06}

\acmConference[CAIR 2017]{The First International Workshop on Conversational Approaches to Information Retrieval}{August 2017}{Tokyo, Japan} 
\acmYear{2017}
\copyrightyear{2017}
\acmPrice{15.00}

\title{Evaluating Personal Assistants on Mobile devices}
\subtitle{Conceptual Paper}

\begin{document}

\author{Julia Kiseleva}
\affiliation{%
  \institution{UserSat.com \& University of Amsterdam}
  \city{Amsterdam} 
  \country{the Netherlands} 
}
\email{j.kiseleva@uva.nl}

\author{Maarten de Rijke}
\affiliation{%
  \institution{University of Amsterdam}
  \city{Amsterdam} 
  \country{the Netherlands} 
}
\email{derijke@uva.nl}

\begin{abstract}
The iPhone was introduced only a decade ago in 2007, but has fundamentally changed the way we interact with online information.  Mobile devices differ radically from classic command-based and point-and-click user interfaces, now allowing for gesture-based interaction using fine-grained touch and swipe signals. 
Due to the rapid growth in the use of voice-controlled intelligent personal assistants on mobile devices, such as  Microsoft's Cortana,  Google Now, and Apple's Siri, mobile devices have become personal, allowing us to be online all the time, and assist us in any task, both in work and in our daily lives, making context a crucial factor to consider.  

Mobile usage is now exceeding desktop usage, and is still growing at a rapid rate, yet our main ways of training and evaluating personal assistants are still based on (and framed in) classical desktop interactions, focusing on explicit queries, clicks, and dwell time spent.
However, modern user interaction with mobile devices is radically different due to touch screens with gesture- and voice-based control and the varying context of use, e.g., in a car, by bike, often invalidating the assumptions underlying today's user satisfaction evaluation.  

There is an urgent need to understand voice- and gesture-based interaction, taking all interaction signals and context into account in appropriate ways. 
We propose a research agenda for developing methods to evaluate and improve \casat with \mi{}s using \gbs at scale.  
\end{abstract}

%
%



\maketitle

\input{1-introduction}

\input{2-research}

\input{3-method}

\bibliographystyle{ACM-Reference-Format}
\bibliography{julia} 

\end{document}

%% file: 1-introduction.tex

\section{Introduction}

Recent years have witnessed an explosive growth in the usage of \device. The usage of mobile phones increased five-fold from 11.78\% in October 2012 to 53.01\% in December 2016,\footnote{\url{http://gs.statcounter.com/\#desktop+mobile+tablet-comparison-ww-monthly-201208-201612}} and it overtook the usage of desktops in October 2016. This spike happens due to availability of personal assistants. Spoken dialogue systems have been thoroughly studied in the literature~\citep{McTear_2002,Tur_2013,Tur_2007,Tur_signal_2013}. However, it has only been in recent years that a new generation of personal assistants, powered by voice, such as Apple's Siri, Microsoft's Cortana, Google Now, have become common and popular on mobile devices.  One of the reasons for the increased adoption is the recent significant improvement in accuracy of automatic speech recognition~\citep{Negri_2014}.

Evaluation of effectiveness is an essential part of developing any interactive system such as web search and e-commerce applications. Modern evaluation methods, which were developed for desktops, heavily rely on interaction data, e.g., explicit queries and clicks that are massively logged~\citep{Chapelle_cikm_2019,Chuklin_sigir_2013,Chuklin_book_2015,Yilmaz_cikm_2010}. However, \is on mobile devices are different due to the context of use and gesture- and voice-based control, like swipes, touch and voice conversations~\citep{kiseleva_sigir_2016, Williams_www_2016,Bachynskyi_CHI_2014, Lehtovirtaa_CHI_2014,WeigelCHI2015,OulasvirtaCHI2005}. 
As a consequence, there is an urgent need to develop new scalable techniques for understanding \casat for gesture- and voice-controlled devices. 

\rqmain
The overall aim leads to three specific research questions:
\begin{itemize}
\item \rqone
\item \rqtwo
\item \rqthree
\end{itemize}

%% file: 2-research.tex

\section{Scientific Challenges}
\label{sec:challenges}

\begin{figure*}[!t]
\centering
\includegraphics[width=.9\textwidth]{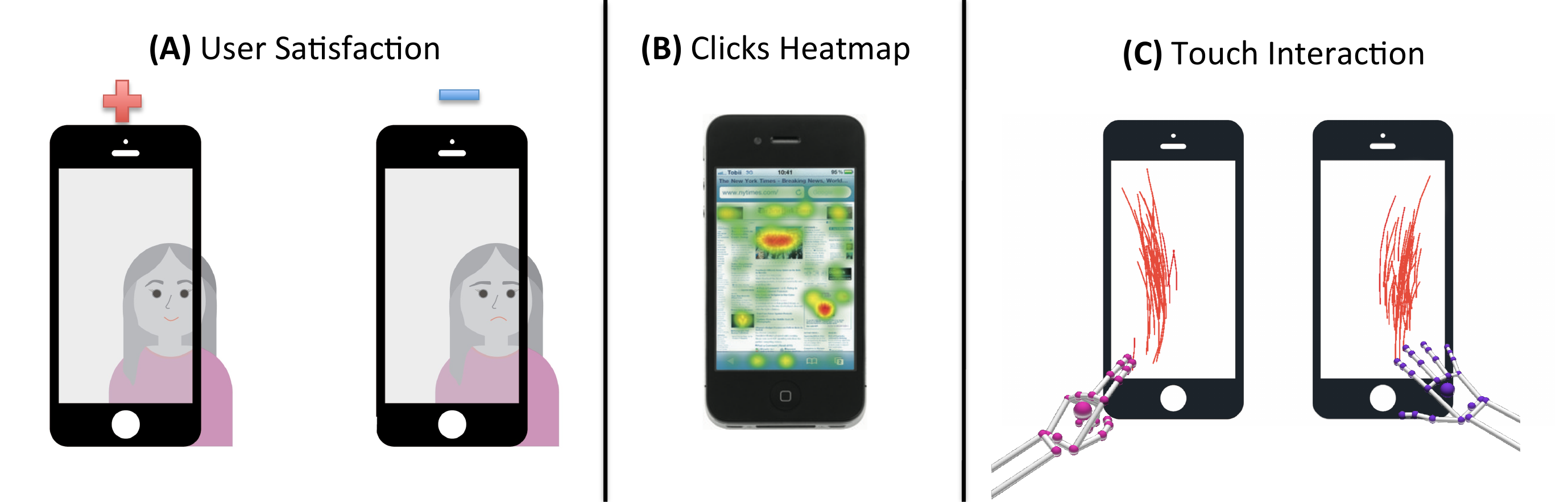}
\caption{Illustrations of (A) how to define user satisfaction, (B) where users click on the mobile screen, and (C) how touches are tracked}
\label{fig:clicks_vs_swipes}
\end{figure*}

We list four central challenges that provide the background for the research questions listed above.

\paragraph{How and why to evaluate the effectiveness of an \ias?}  Previously, a common practice for evaluation was to create a `gold' standard\footnote{A gold standard is a set of ``correct'' answers as judged by editorial judges.}~\citep{sanderson-test-2010}.
In modern \iass, there may be no general ``correct'' answers since the answers are highly personalized and contextualized, e.g., to a user's location~\citep{alves2012,wu_2012,kiseleva_dddm_2013} or a user's past preferences~\citep{Ho_wsdm_2014,Kiseleva_sigir_industry_2015,Xiang_kdd_2010}. 
User satisfaction is widely adopted as a subjective measure of the quality of the search experience~\citep{Kelly_2009}.
Nowadays, it has become common practice to evaluate \iass on desktops by analysing \is such as clicks (if users like they click) and dwell time (the actual length of time that a visitor spends on a page)~\citep{Joachims_2002,Joachims_2005, Fox_trans_2005, Hassan_wsdm_2010, Agichtein_sigir_2006_1, Chapelle_cikm_2019,Chuklin_sigir_2013, Chuklin_book_2015}.

Currently, the research community is facing a challenge to evaluate user satisfaction at scale.
Very large scale online controlled experiments, such as A/B testing and interleaving, have become a widely used technique for controlling and improving search quality based on data-driven decisions~\citep{hofmann-online-2016}. This methodology has been adopted by many leading companies \citep{Deng_www_2014,Tang_kdd_2010,Bakshy_kdd_2013,drutsa_www_2015}. User behavior in voice- and gesture-controlled environment is very different when compared to desktops~\citep{kiseleva_sigir_2016, Williams_www_2016, Lagun_sigir_2015, lagun_cikm_2016,kiseleva_chiir_2016, Williams_sigir_2016}, but our understanding of this difference is still fragmented at best.
Unlike desktop computers with large displays and mouse-keyboard interactions~\cite{GuoCIKM2012, GuoWWW2012, GuoSIGIR2010, Liu2015,Rodden_2008}, personal assistants come on mobile devices that have smaller displays and offer voice commands and a variety of gesture interactions, e.g., touch: swiping and zooming. Moreover, user behavior on mobiles is very context-dependent~\citep{zhu2015mining}.
Therefore, traditional evaluation methods are not applicable for the growing mobile environment.

The fundamental problem limiting current progress in developing \iass for mobile environment is the lack of scalable methods to infer user satisfaction.

\paragraph{Why is context-awareness needed for evaluating \sat?} 
\citet{Kelly_2009} proposes the following definition:~``\emph{satisfaction can be understood as the fulfillment of a specified desire or goal}.'' Online user behavior is \textbf{highly}:
\begin{itemize}
 \item context-dependent~\citep{kiseleva_dddm_2013,kiseleva_expertise_2016,kiselevalpc_www13,Kiseleva_booking_sigir_2016,steffenrendle_sigir_2011,biaoxiang2010, Adomavicius_cars_2010};
 \item sensitive to changes in the outside world~\citep{kiseleva_serp_cikm_2015, kiseleva_cikm_2014}.
\end{itemize}
In a mobile environment, users are dealing with a much richer space of potential contextual situations, e.g., while driving, in the bus, on the way, a slow connection, compared to the relatively static desktop environment.  These conditions have a great impact on `mobile' \sat. 
Similar experiences can be satisfying in one situation (Figure~\ref{fig:clicks_vs_swipes}(A)`+'), e.g., a user is sitting in a hotel lobby with a fast wifi connection, and it can be totally frustrating in another situation (Figure~\ref{fig:clicks_vs_swipes}(A)`--'), e.g., when the same user is driving and having a slow data connection. 

Therefore, situational context has to be studied in far greater details, allowing us to reason about how a user's current environment impacts his satisfaction with personal assistants. 

\paragraph{How to evaluate \casat at scale?}
Eye-track\-ing techniques have been successfully used to gain an initial understanding of user interactions with mobile devices~\citep{lagun_cikm_2016,Lagun_sigir_2014} but they cannot applied at scale. 
In contrast, user gestures and voice commands can be collected and analysed at scale~\citep{Williams_www_2016}. We suggest to model advanced \vgbs to predict \casat for millions of users, which can be easily plugged-in into A/B testing platforms~\citep{Drutsa_wsdm_2015, Bakshy_kdd_2013, kohavi_kdd_2014, Deng_wsdm_2017}.

\paragraph{Why analysing gestures is the way to infer \casat?}
Analysing clicks heat maps, e.g., in Figure~\ref{fig:clicks_vs_swipes}(B), is quite tricky. Because the screen size is small, it is difficult not to click inadvertently, or conversely, to have difficulty clicking an item. Analysing \gesture patters is a better way to infer user satisfaction, as it helps to decipher hidden behavioral aspects, e.g., swipes in Figure~\ref{fig:clicks_vs_swipes}(C) clearly belong to left- and right-handed persons. 
Moreover, touch signals are extremely useful to predict \sat for mobile search~\citep{kiseleva_sigir_2016, Williams_www_2016}.
Movements of the human body, e.g., gestures, reflect emotions~\citep{Bernhardt_thesis_10,Bernhardt_ivic_R09} that are closely connected with user satisfaction (Figure~\ref{fig:clicks_vs_swipes}(A)). 
User emotions are used to evaluate voice-controlled systems~\citep{Koolagudi_2012, Ren_icme_2014}, e.g., changes in user intonation~\citep{Young_2016, Shokouhi_www_2016}.
We propose to exploit \gesture and \voice interaction to infer \casat in mobile environment because they:

\begin{itemize}
\item are the primary ways to interact with mobile devices;
\item are very sensitive to situational and behavioral aspects (Figure~\ref{fig:clicks_vs_swipes}(C));
\item reveal \textsl{user emotions}: satisfaction (Figure~\ref{fig:clicks_vs_swipes}(A)`+') and frustration (Figure~\ref{fig:clicks_vs_swipes}(A)`--');
\item are highly scalable, both w.r.t.\ collection and analysis.
\end{itemize}

%% file: 3-method.tex

\section{Discussion}

We propose ventures into a new area of research, by moving beyond the interaction models effective for desktop applications 
and fully embrace the new paradigm of gesture- and voice-controlled personal assistants on mobiles. This has very large potential impact on billions of users around the world, but also has large scientific impact.  The proposed research will significantly contribute to our scientific understanding of \sat in mobile environment, and the value of massive scale gesture-based interaction logs to infer user satisfaction based on complex and subtle interaction patterns.  Obtaining insight into this value is crucial if we want to evaluate new algorithms of information representations for different type of applications.



\begin{figure}
    \centering
    \includegraphics[width=.4\textwidth]{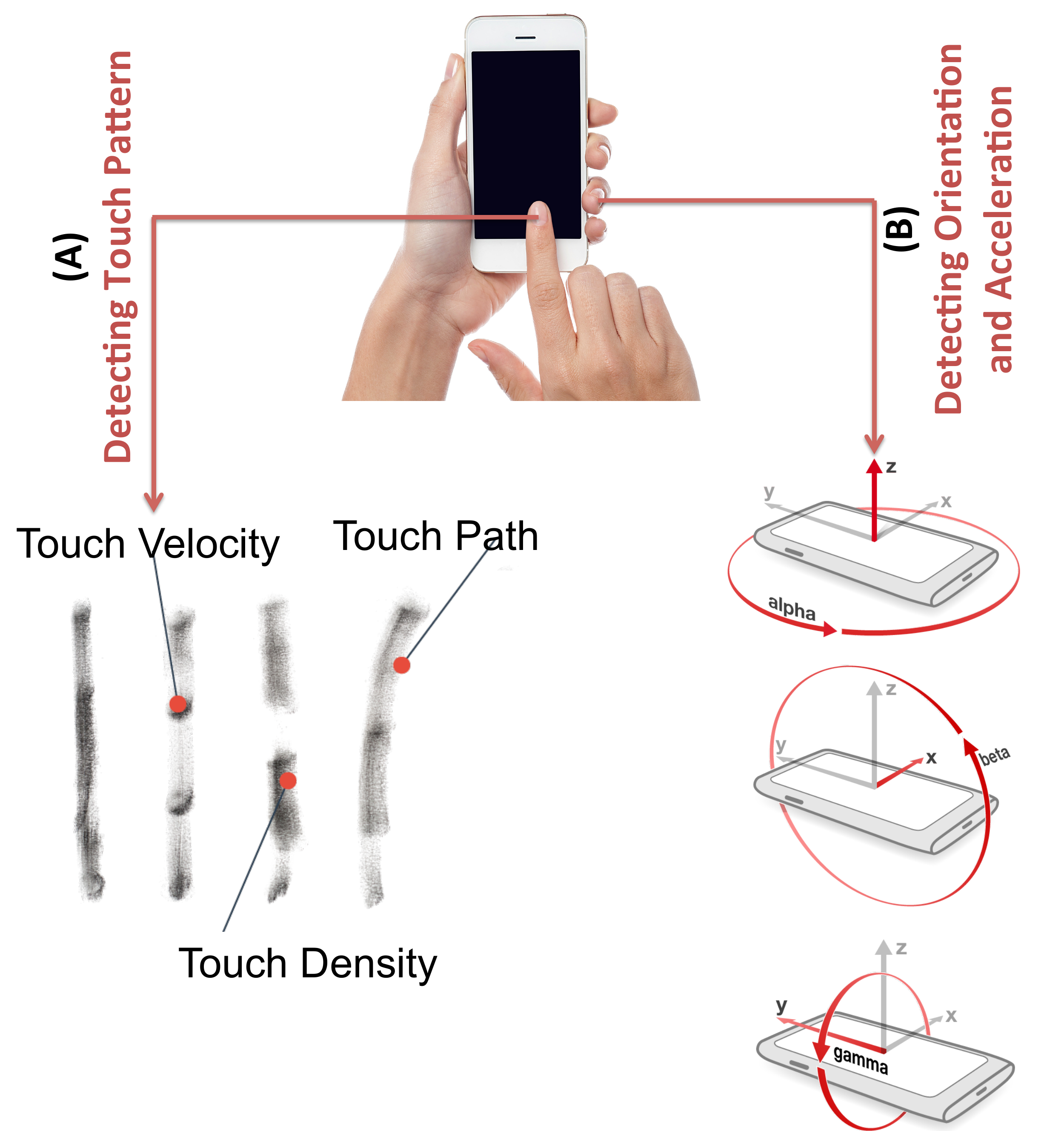}
    \caption{\label{fig:model_gestures} Modelling gesture-based interactions.}
\end{figure}


\paragraph{\emph{\rqone}}
We need to encompass all gesture- and voice-based features related users' interaction with \iass. Capturing \textit{touch events} (Figure~\ref{fig:model_gestures}(A)) is difficult in practice~\citep{Huang_hcir_2012}; however, it is possible to infer touch-based interactions based on the mobile viewport~\citep{Williams_www_2016,kiseleva_sigir_2016}.\footnote{The viewport is the visible region on the device.} For instance, if an element is visible in the viewport at some point in time and then no longer visible, one can infer that a gesture must have taken place. To get a complete view of user gestures, we should capture \begin{inparaenum}[(1)] \item \textit{orientation and acceleration of a device in space} (Figure~\ref{fig:model_gestures}(B)) that will allow us to model users' hands position; \item the GPS signal to infer changes in user locations; \item movement events, e.g. the `shake.'\end{inparaenum}\ We could use this rich set of \gesture features to build an advanced representation of interactions in a mobile setting.

\paragraph{\emph{\rqtwo}} 
As a starting point, we can start from the approach presented in~\citep{kiseleva_sigir_2016} to define user satisfaction with mobile interactions at the session-level. Then, one should extend it by introducing context-aware~\citep{kiselevalpc_www13, kiseleva_dddm_2013, Kiseleva_booking_sigir_2016} and changing environments~\citep{kiseleva_serp_cikm_2015,kiseleva_cikm_2014}.  In addition to unsupervised logs, dedicated user experiments should be conducted to gather rich, explicitly annotated data for further analysis and validation.

\paragraph{\emph{\rqthree}}
Recently, deep neural networks have given rise to significant performance improvements
in speech recognition~\citep{X12a} and computer vision tasks~\citep{lecun2015deep}. They have also led to exciting breakthroughs in novel application areas such as automatic voice translation~\citep{lewis2015skype}, image captioning~\citep{vinyals2015show}, and conversational assistants~\citep{VinyalsL15,GriolC16}. However, there are only few publications on using deep neural networks to model user interaction behavior. So far these have been confined to desktop settings~\citep{borisov2016neural,borisovcontext_sigir_2016, mitra2015exploring}, where the advantage of neural approaches has been clearly demonstrated; the time is right to put them to work in a mobile setting.

%% file: cair2017-ws-julia.bbl

\begin{thebibliography}{00}


\ifx \showCODEN    \undefined \def \showCODEN     #1{\unskip}     \fi
\ifx \showDOI      \undefined \def \showDOI       #1{#1}\fi
\ifx \showISBNx    \undefined \def \showISBNx     #1{\unskip}     \fi
\ifx \showISBNxiii \undefined \def \showISBNxiii  #1{\unskip}     \fi
\ifx \showISSN     \undefined \def \showISSN      #1{\unskip}     \fi
\ifx \showLCCN     \undefined \def \showLCCN      #1{\unskip}     \fi
\ifx \shownote     \undefined \def \shownote      #1{#1}          \fi
\ifx \showarticletitle \undefined \def \showarticletitle #1{#1}   \fi
\ifx \showURL      \undefined \def \showURL       {\relax}        \fi
\providecommand\bibfield[2]{#2}
\providecommand\bibinfo[2]{#2}
\providecommand\natexlab[1]{#1}
\providecommand\showeprint[2][]{arXiv:#2}

\bibitem[\protect\citeauthoryear{Adomavicius and Tuzhilin}{Adomavicius and
  Tuzhilin}{2010}]%
        {Adomavicius_cars_2010}
\bibfield{author}{\bibinfo{person}{Gediminas Adomavicius} {and}
  \bibinfo{person}{Alexander Tuzhilin}.} \bibinfo{year}{2010}\natexlab{}.
\newblock \showarticletitle{Context-Aware Recommender Systems}.
\newblock \bibinfo{journal}{{\em CARS\/}} (\bibinfo{year}{2010}).
\newblock


\bibitem[\protect\citeauthoryear{Agichtein, Brill, and Dumais}{Agichtein
  et~al\mbox{.}}{2006}]%
        {Agichtein_sigir_2006_1}
\bibfield{author}{\bibinfo{person}{Eugene Agichtein}, \bibinfo{person}{Eric
  Brill}, {and} \bibinfo{person}{Susan~T. Dumais}.}
  \bibinfo{year}{2006}\natexlab{}.
\newblock \showarticletitle{Improving web search ranking by incorporating user
  behavior information}. In \bibinfo{booktitle}{{\em SIGIR}}.
  \bibinfo{pages}{19--26}.
\newblock


\bibitem[\protect\citeauthoryear{Alves and C.Pereira}{Alves and
  C.Pereira}{2012}]%
        {alves2012}
\bibfield{author}{\bibinfo{person}{Ana~O. Alves} {and}
  \bibinfo{person}{Fracisco C.Pereira}.} \bibinfo{year}{2012}\natexlab{}.
\newblock \showarticletitle{Making sense of location context}. In
  \bibinfo{booktitle}{{\em ContextDD}}.
\newblock


\bibitem[\protect\citeauthoryear{Bachynskyi, Oulasvirta, Palmas, and
  Weinkauf}{Bachynskyi et~al\mbox{.}}{2014}]%
        {Bachynskyi_CHI_2014}
\bibfield{author}{\bibinfo{person}{Myroslav Bachynskyi}, \bibinfo{person}{Antti
  Oulasvirta}, \bibinfo{person}{Gregorio Palmas}, {and} \bibinfo{person}{Tino
  Weinkauf}.} \bibinfo{year}{2014}\natexlab{}.
\newblock \showarticletitle{Is motion capture-based biomechanical simulation
  valid for HCI studies?: study and implications}. In \bibinfo{booktitle}{{\em
  CHI}}. \bibinfo{pages}{3215--3224}.
\newblock


\bibitem[\protect\citeauthoryear{Bakshy and Eckles}{Bakshy and Eckles}{2013}]%
        {Bakshy_kdd_2013}
\bibfield{author}{\bibinfo{person}{Eytan Bakshy} {and} \bibinfo{person}{Dean
  Eckles}.} \bibinfo{year}{2013}\natexlab{}.
\newblock \showarticletitle{Uncertainty in online experiments with dependent
  data: an evaluation of bootstrap methods}. In \bibinfo{booktitle}{{\em KDD}}.
  \bibinfo{pages}{1303--1311}.
\newblock


\bibitem[\protect\citeauthoryear{Bergstrom-Lehtovirtaa and
  Oulasvirta}{Bergstrom-Lehtovirtaa and Oulasvirta}{2014}]%
        {Lehtovirtaa_CHI_2014}
\bibfield{author}{\bibinfo{person}{Joanna Bergstrom-Lehtovirtaa} {and}
  \bibinfo{person}{Antti Oulasvirta}.} \bibinfo{year}{2014}\natexlab{}.
\newblock \showarticletitle{Modeling the functional area of the thumb on mobile
  touchscreen surfaces}. In \bibinfo{booktitle}{{\em SIGCHI}}.
  \bibinfo{pages}{1991--2000}.
\newblock


\bibitem[\protect\citeauthoryear{Bernhardt}{Bernhardt}{2010}]%
        {Bernhardt_thesis_10}
\bibfield{author}{\bibinfo{person}{Daniel Bernhardt}.}
  \bibinfo{year}{2010}\natexlab{}.
\newblock {\em \bibinfo{title}{Emotion inference from human body motion}}.
\newblock \bibinfo{thesistype}{Ph.D. Dissertation}. \bibinfo{school}{University
  of Cambridge, {UK}}.
\newblock


\bibitem[\protect\citeauthoryear{Bernhardt and Robinson}{Bernhardt and
  Robinson}{2009}]%
        {Bernhardt_ivic_R09}
\bibfield{author}{\bibinfo{person}{Daniel Bernhardt} {and}
  \bibinfo{person}{Peter Robinson}.} \bibinfo{year}{2009}\natexlab{}.
\newblock \showarticletitle{Detecting Emotions from Connected Action
  Sequences}. In \bibinfo{booktitle}{{\em IVIC}}. \bibinfo{pages}{1--11}.
\newblock
\showDOI{%
\url{https://doi.org/10.1007/978-3-642-05036-7_1}}


\bibitem[\protect\citeauthoryear{Borisov, Markov, de~Rijke, and
  Serdyukov}{Borisov et~al\mbox{.}}{2016a}]%
        {borisovcontext_sigir_2016}
\bibfield{author}{\bibinfo{person}{Alexey Borisov}, \bibinfo{person}{Ilya
  Markov}, \bibinfo{person}{Maarten de Rijke}, {and} \bibinfo{person}{Pavel
  Serdyukov}.} \bibinfo{year}{2016}\natexlab{a}.
\newblock \showarticletitle{A Context-aware Time Model for Web Search}. In
  \bibinfo{booktitle}{{\em SIGIR}}. \bibinfo{pages}{205--214}.
\newblock


\bibitem[\protect\citeauthoryear{Borisov, Markov, de~Rijke, and
  Serdyukov}{Borisov et~al\mbox{.}}{2016b}]%
        {borisov2016neural}
\bibfield{author}{\bibinfo{person}{Alexey Borisov}, \bibinfo{person}{Ilya
  Markov}, \bibinfo{person}{Maarten de Rijke}, {and} \bibinfo{person}{Pavel
  Serdyukov}.} \bibinfo{year}{2016}\natexlab{b}.
\newblock \showarticletitle{A Neural Click Model for Web Search}. In
  \bibinfo{booktitle}{{\em WWW}}. International World Wide Web Conferences
  Steering Committee, \bibinfo{pages}{531--541}.
\newblock


\bibitem[\protect\citeauthoryear{Chapelle, Metlzer, Zhang, and
  Grinspan}{Chapelle et~al\mbox{.}}{2010}]%
        {Chapelle_cikm_2019}
\bibfield{author}{\bibinfo{person}{Olivier Chapelle}, \bibinfo{person}{Donald
  Metlzer}, \bibinfo{person}{Ya Zhang}, {and} \bibinfo{person}{Pierre
  Grinspan}.} \bibinfo{year}{2010}\natexlab{}.
\newblock \showarticletitle{Expected reciprocal rank for graded relevance}. In
  \bibinfo{booktitle}{{\em CIKM}}. \bibinfo{pages}{621--630}.
\newblock


\bibitem[\protect\citeauthoryear{Chuklin, Markov, and de~Rijke}{Chuklin
  et~al\mbox{.}}{2015}]%
        {Chuklin_book_2015}
\bibfield{author}{\bibinfo{person}{Aleksandr Chuklin}, \bibinfo{person}{Ilya
  Markov}, {and} \bibinfo{person}{Maarten de Rijke}.}
  \bibinfo{year}{2015}\natexlab{}.
\newblock \bibinfo{booktitle}{{\em Click Models for Web Search}}.
\newblock \bibinfo{publisher}{Morgan \& Claypool Publishers}.
\newblock
\showISBNx{1627056475, 9781627056472}


\bibitem[\protect\citeauthoryear{Chuklin, Serdyukov, and de~Rijke}{Chuklin
  et~al\mbox{.}}{2013}]%
        {Chuklin_sigir_2013}
\bibfield{author}{\bibinfo{person}{Aleksandr Chuklin}, \bibinfo{person}{Pavel
  Serdyukov}, {and} \bibinfo{person}{Maarten de Rijke}.}
  \bibinfo{year}{2013}\natexlab{}.
\newblock \showarticletitle{Click model-based information retrieval metrics}.
  In \bibinfo{booktitle}{{\em SIGIR}}. \bibinfo{pages}{493--502}.
\newblock


\bibitem[\protect\citeauthoryear{Deng, Li, and Guo}{Deng et~al\mbox{.}}{2014}]%
        {Deng_www_2014}
\bibfield{author}{\bibinfo{person}{Alex Deng}, \bibinfo{person}{Tianxi Li},
  {and} \bibinfo{person}{Yu Guo}.} \bibinfo{year}{2014}\natexlab{}.
\newblock \showarticletitle{Statistical inference in two-stage online
  controlled experiments with treatment selection and validation.}. In
  \bibinfo{booktitle}{{\em WWW}}. \bibinfo{pages}{609--618}.
\newblock


\bibitem[\protect\citeauthoryear{Deng, Lu, and Litz}{Deng
  et~al\mbox{.}}{2017}]%
        {Deng_wsdm_2017}
\bibfield{author}{\bibinfo{person}{Alex Deng}, \bibinfo{person}{Jiannan Lu},
  {and} \bibinfo{person}{Jonathan Litz}.} \bibinfo{year}{2017}\natexlab{}.
\newblock \showarticletitle{Trustworthy analysis of online A/B tests: Pitfalls,
  challenges and solutions}. In \bibinfo{booktitle}{{\em WSDM}}.
\newblock


\bibitem[\protect\citeauthoryear{Drutsa, Gusev, and Serdyukov}{Drutsa
  et~al\mbox{.}}{2015a}]%
        {Drutsa_wsdm_2015}
\bibfield{author}{\bibinfo{person}{Alexey Drutsa}, \bibinfo{person}{Gleb
  Gusev}, {and} \bibinfo{person}{Pavel Serdyukov}.}
  \bibinfo{year}{2015}\natexlab{a}.
\newblock \showarticletitle{Engagement Periodicity in Search Engine Usage:
  Analysis and its Application to Search Quality Evaluation}. In
  \bibinfo{booktitle}{{\em WSDM}}. \bibinfo{pages}{27--36}.
\newblock


\bibitem[\protect\citeauthoryear{Drutsa, Gusev, and Serdyukov}{Drutsa
  et~al\mbox{.}}{2015b}]%
        {drutsa_www_2015}
\bibfield{author}{\bibinfo{person}{Alexey Drutsa}, \bibinfo{person}{Gleb
  Gusev}, {and} \bibinfo{person}{Pavel Serdyukov}.}
  \bibinfo{year}{2015}\natexlab{b}.
\newblock \showarticletitle{Future User Engagement Prediction and Its
  Application to Improve the Sensitivity of Online Experiments}. In
  \bibinfo{booktitle}{{\em WWW}}. \bibinfo{pages}{256--266}.
\newblock


\bibitem[\protect\citeauthoryear{Fox, Karnawat, Mydland, Dumais, and White}{Fox
  et~al\mbox{.}}{2005}]%
        {Fox_trans_2005}
\bibfield{author}{\bibinfo{person}{Steve Fox}, \bibinfo{person}{Kuldeep
  Karnawat}, \bibinfo{person}{Mark Mydland}, \bibinfo{person}{Susan~T. Dumais},
  {and} \bibinfo{person}{Thomas White}.} \bibinfo{year}{2005}\natexlab{}.
\newblock \showarticletitle{Evaluating implicit measures to improve web
  search}.
\newblock \bibinfo{journal}{{\em TOIS\/}} \bibinfo{volume}{23},
  \bibinfo{number}{2} (\bibinfo{year}{2005}), \bibinfo{pages}{147--168}.
\newblock


\bibitem[\protect\citeauthoryear{Griol and Callejas}{Griol and
  Callejas}{2016}]%
        {GriolC16}
\bibfield{author}{\bibinfo{person}{David Griol} {and} \bibinfo{person}{Zoraida
  Callejas}.} \bibinfo{year}{2016}\natexlab{}.
\newblock \showarticletitle{A Neural Network Approach to Intention Modeling for
  User-Adapted Conversational Agents}.
\newblock \bibinfo{journal}{{\em Comp. Int. and Neurosc.\/}}
  \bibinfo{volume}{2016} (\bibinfo{year}{2016}),
  \bibinfo{pages}{8402127:1--8402127:11}.
\newblock


\bibitem[\protect\citeauthoryear{Guo and Agichtein}{Guo and Agichtein}{2010}]%
        {GuoSIGIR2010}
\bibfield{author}{\bibinfo{person}{Qi Guo} {and} \bibinfo{person}{Eugene
  Agichtein}.} \bibinfo{year}{2010}\natexlab{}.
\newblock \showarticletitle{Ready to buy or just browsing? Detecting Web
  Searcher Goals from Interaction Data}. In \bibinfo{booktitle}{{\em SIGIR}}.
  \bibinfo{pages}{130--137}.
\newblock


\bibitem[\protect\citeauthoryear{Guo and Agichtein}{Guo and Agichtein}{2012}]%
        {GuoWWW2012}
\bibfield{author}{\bibinfo{person}{Qi Guo} {and} \bibinfo{person}{Eugene
  Agichtein}.} \bibinfo{year}{2012}\natexlab{}.
\newblock \showarticletitle{Beyond Dwell Time: Estimating Document Relevance
  from Cursor Movements and other Post-click Searcher Behavior}. In
  \bibinfo{booktitle}{{\em WWW}}. \bibinfo{publisher}{ACM Press},
  \bibinfo{address}{New York, New York, USA}, \bibinfo{pages}{569--578}.
\newblock


\bibitem[\protect\citeauthoryear{Guo, Lagun, and Agichtein}{Guo
  et~al\mbox{.}}{2012}]%
        {GuoCIKM2012}
\bibfield{author}{\bibinfo{person}{Qi Guo}, \bibinfo{person}{Dmitry Lagun},
  {and} \bibinfo{person}{Eugene Agichtein}.} \bibinfo{year}{2012}\natexlab{}.
\newblock \showarticletitle{Predicting web search success with fine-grained
  interaction data}. In \bibinfo{booktitle}{{\em CIKM}}.
  \bibinfo{pages}{2050--2054}.
\newblock


\bibitem[\protect\citeauthoryear{Hassan, Jones, and Klinkner}{Hassan
  et~al\mbox{.}}{2010}]%
        {Hassan_wsdm_2010}
\bibfield{author}{\bibinfo{person}{Ahmed Hassan}, \bibinfo{person}{Rosie
  Jones}, {and} \bibinfo{person}{Kristina~Lisa Klinkner}.}
  \bibinfo{year}{2010}\natexlab{}.
\newblock \showarticletitle{Beyond {DCG}: User Behavior as a Predictor of a
  Successful Search}. In \bibinfo{booktitle}{{\em WSDM}}.
  \bibinfo{pages}{221--230}.
\newblock


\bibitem[\protect\citeauthoryear{Hinton, Deng, Yu, Dahl, Mohamed, Jaitly,
  Senior, Vanhoucke, Nguyen, Sainath, et~al\mbox{.}}{Hinton
  et~al\mbox{.}}{2012}]%
        {X12a}
\bibfield{author}{\bibinfo{person}{Geoffrey Hinton}, \bibinfo{person}{Li Deng},
  \bibinfo{person}{Dong Yu}, \bibinfo{person}{George~E Dahl},
  \bibinfo{person}{Abdel-rahman Mohamed}, \bibinfo{person}{Navdeep Jaitly},
  \bibinfo{person}{Andrew Senior}, \bibinfo{person}{Vincent Vanhoucke},
  \bibinfo{person}{Patrick Nguyen}, \bibinfo{person}{Tara~N Sainath},
  {et~al\mbox{.}}} \bibinfo{year}{2012}\natexlab{}.
\newblock \showarticletitle{Deep Neural Networks for Acoustic Modeling in
  Speech Recognition: The Shared Views of Four Research Groups}.
\newblock \bibinfo{journal}{{\em {IEEE} Signal Process. Mag.\/}}
  \bibinfo{volume}{29}, \bibinfo{number}{6} (\bibinfo{year}{2012}),
  \bibinfo{pages}{82--97}.
\newblock


\bibitem[\protect\citeauthoryear{Ho, Chiang, and Hsu Yung-Jen}{Ho
  et~al\mbox{.}}{2014}]%
        {Ho_wsdm_2014}
\bibfield{author}{\bibinfo{person}{Yu-Chieh Ho}, \bibinfo{person}{Yi-Ting
  Chiang}, {and} \bibinfo{person}{Jane Hsu Yung-Jen}.}
  \bibinfo{year}{2014}\natexlab{}.
\newblock \showarticletitle{Who likes it more?: mining worth-recommending items
  from long tails by modeling relative preference}. In \bibinfo{booktitle}{{\em
  WSDM}}. \bibinfo{pages}{253--262}.
\newblock


\bibitem[\protect\citeauthoryear{Hofmann, Li, and Radlinski}{Hofmann
  et~al\mbox{.}}{2016}]%
        {hofmann-online-2016}
\bibfield{author}{\bibinfo{person}{Katja Hofmann}, \bibinfo{person}{Lihong Li},
  {and} \bibinfo{person}{Filip Radlinski}.} \bibinfo{year}{2016}\natexlab{}.
\newblock \showarticletitle{Online Evaluation for Information Retrieval}.
\newblock \bibinfo{journal}{{\em Foundations and Trends in Information
  Retrieval\/}} \bibinfo{volume}{10}, \bibinfo{number}{1}
  (\bibinfo{year}{2016}), \bibinfo{pages}{1--117}.
\newblock


\bibitem[\protect\citeauthoryear{Huang and Diriye}{Huang and Diriye}{2012}]%
        {Huang_hcir_2012}
\bibfield{author}{\bibinfo{person}{Jeff Huang} {and} \bibinfo{person}{Abdigani
  Diriye}.} \bibinfo{year}{2012}\natexlab{}.
\newblock \showarticletitle{Web User Interaction Mining From Touch Enabled
  Mobile Devices}. In \bibinfo{booktitle}{{\em HCIR Workshop}}.
\newblock


\bibitem[\protect\citeauthoryear{Joachims}{Joachims}{2002}]%
        {Joachims_2002}
\bibfield{author}{\bibinfo{person}{Thorsten Joachims}.}
  \bibinfo{year}{2002}\natexlab{}.
\newblock \showarticletitle{Optimizing search engines using clickthrough data}.
  In \bibinfo{booktitle}{{\em KDD}}. \bibinfo{pages}{133--142}.
\newblock


\bibitem[\protect\citeauthoryear{Joachims, Granka, Pan, Hembrooke, and
  Gay}{Joachims et~al\mbox{.}}{2005}]%
        {Joachims_2005}
\bibfield{author}{\bibinfo{person}{Thorsten Joachims}, \bibinfo{person}{Laura
  Granka}, \bibinfo{person}{Bing Pan}, \bibinfo{person}{Helene Hembrooke},
  {and} \bibinfo{person}{Geri Gay}.} \bibinfo{year}{2005}\natexlab{}.
\newblock \showarticletitle{Accurately interpreting clickthrough data as
  implicit feedback}. In \bibinfo{booktitle}{{\em SIGIR}}.
  \bibinfo{pages}{154--161}.
\newblock


\bibitem[\protect\citeauthoryear{Kelly}{Kelly}{2009}]%
        {Kelly_2009}
\bibfield{author}{\bibinfo{person}{Diane Kelly}.}
  \bibinfo{year}{2009}\natexlab{}.
\newblock \showarticletitle{Methods for evaluating interactive information
  retrieval systems with users}.
\newblock \bibinfo{journal}{{\em FTIR\/}} \bibinfo{volume}{3},
  \bibinfo{number}{1-2} (\bibinfo{year}{2009}), \bibinfo{pages}{1--224}.
\newblock


\bibitem[\protect\citeauthoryear{Kiseleva, Crestan, Brigo, and Dittel}{Kiseleva
  et~al\mbox{.}}{2014}]%
        {kiseleva_cikm_2014}
\bibfield{author}{\bibinfo{person}{Julia Kiseleva}, \bibinfo{person}{Eric
  Crestan}, \bibinfo{person}{Riccardo Brigo}, {and} \bibinfo{person}{Roland
  Dittel}.} \bibinfo{year}{2014}\natexlab{}.
\newblock \showarticletitle{Modelling and Detecting Changes in User
  Satisfaction}. In \bibinfo{booktitle}{{\em CIKM}}.
  \bibinfo{pages}{1449--1458}.
\newblock


\bibitem[\protect\citeauthoryear{Kiseleva, Kamps, Nikulin, and
  Makarov}{Kiseleva et~al\mbox{.}}{2015}]%
        {kiseleva_serp_cikm_2015}
\bibfield{author}{\bibinfo{person}{Julia Kiseleva}, \bibinfo{person}{Jaap
  Kamps}, \bibinfo{person}{Vadim Nikulin}, {and} \bibinfo{person}{Nikita
  Makarov}.} \bibinfo{year}{2015}\natexlab{}.
\newblock \showarticletitle{Behavioral Dynamics from the {SERP}'s Perspective:
  What are failed {SERP}s and how to fix them?}. In \bibinfo{booktitle}{{\em
  CIKM}}. \bibinfo{pages}{1561--1570}.
\newblock


\bibitem[\protect\citeauthoryear{Kiseleva, Lam, Pechenizkiy, and
  Calders}{Kiseleva et~al\mbox{.}}{2013a}]%
        {kiselevalpc_www13}
\bibfield{author}{\bibinfo{person}{Julia Kiseleva},
  \bibinfo{person}{Hoang~Thanh Lam}, \bibinfo{person}{Mykola Pechenizkiy},
  {and} \bibinfo{person}{Toon Calders}.} \bibinfo{year}{2013}\natexlab{a}.
\newblock \showarticletitle{Discovering Temporal Hidden Contexts in Web
  Sessions for User Trail Prediction}. In \bibinfo{booktitle}{{\em WWW Workshop
  TempWeb}}. \bibinfo{pages}{1067--1074}.
\newblock


\bibitem[\protect\citeauthoryear{Kiseleva, Lam, Pechenizkiy, and
  Calders}{Kiseleva et~al\mbox{.}}{2013b}]%
        {kiseleva_dddm_2013}
\bibfield{author}{\bibinfo{person}{Julia Kiseleva},
  \bibinfo{person}{Hoang~Thanh Lam}, \bibinfo{person}{Mykola Pechenizkiy},
  {and} \bibinfo{person}{Toon Calders}.} \bibinfo{year}{2013}\natexlab{b}.
\newblock \showarticletitle{Predicting Current User Intent with Contextual
  Markov Models}. In \bibinfo{booktitle}{{\em ICDMW}}.
  \bibinfo{pages}{391--398}.
\newblock


\bibitem[\protect\citeauthoryear{Kiseleva, {Montes Garc{\'\i}a}, Kamps, and
  Spirin}{Kiseleva et~al\mbox{.}}{2016}]%
        {kiseleva_expertise_2016}
\bibfield{author}{\bibinfo{person}{Julia Kiseleva}, \bibinfo{person}{Alejandro
  {Montes Garc{\'\i}a}}, \bibinfo{person}{Jaap Kamps}, {and}
  \bibinfo{person}{Nikita Spirin}.} \bibinfo{year}{2016}\natexlab{}.
\newblock \showarticletitle{The Impact of Technical Domain Expertise on Search
  Behavior and Task Outcome}. In \bibinfo{booktitle}{{\em Proceedings of WSDM
  Workshop QRUMS}}.
\newblock


\bibitem[\protect\citeauthoryear{Kiseleva, Müller, Bernardi, Davis, Kovacek,
  {Stafseng Einarsen}, Kamps, Tuzhilin, and Hiemstra}{Kiseleva
  et~al\mbox{.}}{2015}]%
        {Kiseleva_sigir_industry_2015}
\bibfield{author}{\bibinfo{person}{Julia Kiseleva}, \bibinfo{person}{Melanie
  J.~I. Müller}, \bibinfo{person}{Lucas Bernardi}, \bibinfo{person}{Chad
  Davis}, \bibinfo{person}{Ivan Kovacek}, \bibinfo{person}{Mats {Stafseng
  Einarsen}}, \bibinfo{person}{Jaap Kamps}, \bibinfo{person}{Alexander
  Tuzhilin}, {and} \bibinfo{person}{Djoerd Hiemstra}.}
  \bibinfo{year}{2015}\natexlab{}.
\newblock \showarticletitle{Where to Go on Your Next Trip? Optimizing Travel
  Destinations Based on User Preferences}. In \bibinfo{booktitle}{{\em SIGIR}}.
  \bibinfo{pages}{1097--1100}.
\newblock


\bibitem[\protect\citeauthoryear{Kiseleva, Tuzhilin, Kamps, M{\"u}ller,
  Bernardi, Davis, Kovacek, {Stafseng Einarsen}, and Hiemstra}{Kiseleva
  et~al\mbox{.}}{2016a}]%
        {Kiseleva_booking_sigir_2016}
\bibfield{author}{\bibinfo{person}{Julia Kiseleva}, \bibinfo{person}{Alexander
  Tuzhilin}, \bibinfo{person}{Jaap Kamps}, \bibinfo{person}{Melanie J.~I.
  M{\"u}ller}, \bibinfo{person}{Lucas Bernardi}, \bibinfo{person}{Chad Davis},
  \bibinfo{person}{Ivan Kovacek}, \bibinfo{person}{Mats {Stafseng Einarsen}},
  {and} \bibinfo{person}{Djoerd Hiemstra}.} \bibinfo{year}{2016}\natexlab{a}.
\newblock \showarticletitle{Beyond Movie Recommendations: Solving the
  Continuous Cold Start Problem in E-commerceRecommendations}.
\newblock \bibinfo{journal}{{\em CoRR\/}}  \bibinfo{volume}{1607.07904}
  (\bibinfo{year}{2016}).
\newblock


\bibitem[\protect\citeauthoryear{Kiseleva, Williams, Awadallah, Zitouni, Crook,
  and Anastasakos}{Kiseleva et~al\mbox{.}}{2016b}]%
        {kiseleva_sigir_2016}
\bibfield{author}{\bibinfo{person}{Julia Kiseleva}, \bibinfo{person}{Kyle
  Williams}, \bibinfo{person}{Ahmed~Hassan Awadallah}, \bibinfo{person}{Imed
  Zitouni}, \bibinfo{person}{Aidan Crook}, {and} \bibinfo{person}{Tasos
  Anastasakos}.} \bibinfo{year}{2016}\natexlab{b}.
\newblock \showarticletitle{Predicting User Satisfaction with Intelligent
  Assistants}. In \bibinfo{booktitle}{{\em SIGIR}}.
\newblock


\bibitem[\protect\citeauthoryear{Kiseleva, Williams, Jiang, Awadallah, Zitouni,
  Crook, and Anastasakos}{Kiseleva et~al\mbox{.}}{2016c}]%
        {kiseleva_chiir_2016}
\bibfield{author}{\bibinfo{person}{Julia Kiseleva}, \bibinfo{person}{Kyle
  Williams}, \bibinfo{person}{Jiepu Jiang}, \bibinfo{person}{Ahmed~Hassan
  Awadallah}, \bibinfo{person}{Imed Zitouni}, \bibinfo{person}{Aidan Crook},
  {and} \bibinfo{person}{Tasos Anastasakos}.} \bibinfo{year}{2016}\natexlab{c}.
\newblock \showarticletitle{Understanding User Satisfaction with Intelligent
  Assistants}. In \bibinfo{booktitle}{{\em CHIIR}}. \bibinfo{pages}{121 --
  130}.
\newblock


\bibitem[\protect\citeauthoryear{Kohavi, Deng, Longbotham, and Xu}{Kohavi
  et~al\mbox{.}}{2014}]%
        {kohavi_kdd_2014}
\bibfield{author}{\bibinfo{person}{Ron Kohavi}, \bibinfo{person}{Alex Deng},
  \bibinfo{person}{Roger Longbotham}, {and} \bibinfo{person}{Ya Xu}.}
  \bibinfo{year}{2014}\natexlab{}.
\newblock \showarticletitle{Seven rules of thumb for web site experimenters}.
  In \bibinfo{booktitle}{{\em KDD}}. \bibinfo{pages}{1857--1866}.
\newblock


\bibitem[\protect\citeauthoryear{Koolagudi and Rao}{Koolagudi and Rao}{2012}]%
        {Koolagudi_2012}
\bibfield{author}{\bibinfo{person}{Shashidhar~G Koolagudi} {and}
  \bibinfo{person}{K~Sreenivasa Rao}.} \bibinfo{year}{2012}\natexlab{}.
\newblock \showarticletitle{Emotion recognition from speech: a review}.
\newblock \bibinfo{journal}{{\em International journal of speech technology\/}}
  \bibinfo{volume}{15}, \bibinfo{number}{2} (\bibinfo{year}{2012}),
  \bibinfo{pages}{99--117}.
\newblock


\bibitem[\protect\citeauthoryear{Lagun and Agichtein}{Lagun and
  Agichtein}{2015}]%
        {Lagun_sigir_2015}
\bibfield{author}{\bibinfo{person}{Dmitry Lagun} {and} \bibinfo{person}{Eugene
  Agichtein}.} \bibinfo{year}{2015}\natexlab{}.
\newblock \showarticletitle{Inferring Searcher Attention by Jointly Modeling
  User Interactions and Content Salience}. In \bibinfo{booktitle}{{\em SIGIR}}.
  \bibinfo{pages}{483--492}.
\newblock


\bibitem[\protect\citeauthoryear{Lagun, Hsieh, Webster, and Navalpakkam}{Lagun
  et~al\mbox{.}}{2014}]%
        {Lagun_sigir_2014}
\bibfield{author}{\bibinfo{person}{Dmitry Lagun}, \bibinfo{person}{Chih-Hung
  Hsieh}, \bibinfo{person}{Dale Webster}, {and} \bibinfo{person}{Vidhya
  Navalpakkam}.} \bibinfo{year}{2014}\natexlab{}.
\newblock \showarticletitle{Towards better measurement of attention and
  satisfaction in mobile search}. In \bibinfo{booktitle}{{\em SIGIR}}.
  \bibinfo{pages}{113--122}.
\newblock


\bibitem[\protect\citeauthoryear{Lagun, McMahon, and Navalpakkam}{Lagun
  et~al\mbox{.}}{2016}]%
        {lagun_cikm_2016}
\bibfield{author}{\bibinfo{person}{Dmitry Lagun}, \bibinfo{person}{Donal
  McMahon}, {and} \bibinfo{person}{Vidhya Navalpakkam}.}
  \bibinfo{year}{2016}\natexlab{}.
\newblock \showarticletitle{Understanding Mobile Searcher Attention with Rich
  Ad Formats}. In \bibinfo{booktitle}{{\em CIKM}}. \bibinfo{pages}{599--608}.
\newblock


\bibitem[\protect\citeauthoryear{LeCun, Bengio, and Hinton}{LeCun
  et~al\mbox{.}}{2015}]%
        {lecun2015deep}
\bibfield{author}{\bibinfo{person}{Yann LeCun}, \bibinfo{person}{Yoshua
  Bengio}, {and} \bibinfo{person}{Geoffrey Hinton}.}
  \bibinfo{year}{2015}\natexlab{}.
\newblock \showarticletitle{Deep learning}.
\newblock \bibinfo{journal}{{\em Nature\/}} \bibinfo{volume}{521},
  \bibinfo{number}{7553} (\bibinfo{year}{2015}), \bibinfo{pages}{436--444}.
\newblock


\bibitem[\protect\citeauthoryear{Lewis}{Lewis}{2015}]%
        {lewis2015skype}
\bibfield{author}{\bibinfo{person}{William~D Lewis}.}
  \bibinfo{year}{2015}\natexlab{}.
\newblock \showarticletitle{Skype translator: Breaking down language and
  hearing barriers}.
\newblock \bibinfo{journal}{{\em Proceedings of Translating and the Computer
  (TC37)\/}} (\bibinfo{year}{2015}).
\newblock


\bibitem[\protect\citeauthoryear{Liu, Chen, Tang, Sun, Zhang, Ma, and Zhu}{Liu
  et~al\mbox{.}}{2015}]%
        {Liu2015}
\bibfield{author}{\bibinfo{person}{Yiqun Liu}, \bibinfo{person}{Ye Chen},
  \bibinfo{person}{Jinhui Tang}, \bibinfo{person}{Jiashen Sun},
  \bibinfo{person}{Min Zhang}, \bibinfo{person}{Shaoping Ma}, {and}
  \bibinfo{person}{Xuan Zhu}.} \bibinfo{year}{2015}\natexlab{}.
\newblock \showarticletitle{Different Users, Different Opinions: Predicting
  Search Satisfaction with Mouse Movement Information}. In
  \bibinfo{booktitle}{{\em SIGIR}}. \bibinfo{pages}{493--502}.
\newblock


\bibitem[\protect\citeauthoryear{McTear}{McTear}{2002}]%
        {McTear_2002}
\bibfield{author}{\bibinfo{person}{Michael~F. McTear}.}
  \bibinfo{year}{2002}\natexlab{}.
\newblock \showarticletitle{Spoken dialogue technology: enabling the
  conversational user interface}.
\newblock \bibinfo{journal}{{\em ACM Computing Surveys (CSUR)\/}}
  \bibinfo{volume}{34}, \bibinfo{number}{1} (\bibinfo{year}{2002}),
  \bibinfo{pages}{90--169}.
\newblock


\bibitem[\protect\citeauthoryear{Mitra}{Mitra}{2015}]%
        {mitra2015exploring}
\bibfield{author}{\bibinfo{person}{Bhaskar Mitra}.}
  \bibinfo{year}{2015}\natexlab{}.
\newblock \showarticletitle{Exploring session context using distributed
  representations of queries and reformulations}. In \bibinfo{booktitle}{{\em
  SIGIR}}. ACM, \bibinfo{pages}{3--12}.
\newblock


\bibitem[\protect\citeauthoryear{Negri, Turchi, de~Souza, and Falavigna}{Negri
  et~al\mbox{.}}{2014}]%
        {Negri_2014}
\bibfield{author}{\bibinfo{person}{Matteo Negri}, \bibinfo{person}{Marco
  Turchi}, \bibinfo{person}{Jos{\'e} G.~C. de Souza}, {and}
  \bibinfo{person}{Daniele Falavigna}.} \bibinfo{year}{2014}\natexlab{}.
\newblock \showarticletitle{Quality Estimation for Automatic Speech
  Recognition}. In \bibinfo{booktitle}{{\em COLING}}.
  \bibinfo{pages}{1813--1823}.
\newblock


\bibitem[\protect\citeauthoryear{Oulasvirta, Tamminen, Roto, and
  Kuorelahti}{Oulasvirta et~al\mbox{.}}{2005}]%
        {OulasvirtaCHI2005}
\bibfield{author}{\bibinfo{person}{Antti Oulasvirta}, \bibinfo{person}{Sakari
  Tamminen}, \bibinfo{person}{Virpi Roto}, {and} \bibinfo{person}{Jaana
  Kuorelahti}.} \bibinfo{year}{2005}\natexlab{}.
\newblock \showarticletitle{Interaction in 4-second bursts: the fragmented
  nature of attentional resources in mobile HCI}. In \bibinfo{booktitle}{{\em
  SIGCHI}}. \bibinfo{pages}{919--928}.
\newblock


\bibitem[\protect\citeauthoryear{Ren, Jia, Guo, Zhang, and Cai}{Ren
  et~al\mbox{.}}{2014}]%
        {Ren_icme_2014}
\bibfield{author}{\bibinfo{person}{Zhu Ren}, \bibinfo{person}{Jia Jia},
  \bibinfo{person}{Quan Guo}, \bibinfo{person}{Kuo Zhang}, {and}
  \bibinfo{person}{Lianhong Cai}.} \bibinfo{year}{2014}\natexlab{}.
\newblock \showarticletitle{Acoustics, content and geo-information based
  sentiment prediction from large-scale networked voice data}. In
  \bibinfo{booktitle}{{\em ICME}}.
\newblock


\bibitem[\protect\citeauthoryear{Rendle, Gantner, Freudenthaler, and
  Schmidt-Thieme}{Rendle et~al\mbox{.}}{2011}]%
        {steffenrendle_sigir_2011}
\bibfield{author}{\bibinfo{person}{Steffen Rendle}, \bibinfo{person}{Zeno
  Gantner}, \bibinfo{person}{Christoph Freudenthaler}, {and}
  \bibinfo{person}{Lars Schmidt-Thieme}.} \bibinfo{year}{2011}\natexlab{}.
\newblock \showarticletitle{Fast Context-aware Recommendations with
  Factorization Machines}. In \bibinfo{booktitle}{{\em SIGIR}}.
  \bibinfo{pages}{635--644}.
\newblock


\bibitem[\protect\citeauthoryear{Rodden, Fu, Aula, and Spiro}{Rodden
  et~al\mbox{.}}{2008}]%
        {Rodden_2008}
\bibfield{author}{\bibinfo{person}{Kerry Rodden}, \bibinfo{person}{Xin Fu},
  \bibinfo{person}{Anne Aula}, {and} \bibinfo{person}{Ian Spiro}.}
  \bibinfo{year}{2008}\natexlab{}.
\newblock \showarticletitle{Eye-mouse coordination patterns on web search
  results pages}. In \bibinfo{booktitle}{{\em CHI}}.
  \bibinfo{pages}{2997--3002}.
\newblock


\bibitem[\protect\citeauthoryear{Sanderson}{Sanderson}{2010}]%
        {sanderson-test-2010}
\bibfield{author}{\bibinfo{person}{Mark Sanderson}.}
  \bibinfo{year}{2010}\natexlab{}.
\newblock \showarticletitle{Test Collection Based Evaluation of Information
  Retrieval Systems}.
\newblock \bibinfo{journal}{{\em Foundations and Trends in Information
  Retrieval\/}} \bibinfo{volume}{4}, \bibinfo{number}{4}
  (\bibinfo{year}{2010}), \bibinfo{pages}{247--375}.
\newblock


\bibitem[\protect\citeauthoryear{Shokouhi, Ozertem, and Craswell}{Shokouhi
  et~al\mbox{.}}{2016}]%
        {Shokouhi_www_2016}
\bibfield{author}{\bibinfo{person}{Milad Shokouhi}, \bibinfo{person}{Umut
  Ozertem}, {and} \bibinfo{person}{Nick Craswell}.}
  \bibinfo{year}{2016}\natexlab{}.
\newblock \showarticletitle{Did You Say U2 or YouTube? Inferring Implicit
  Transcripts from Voice Search Logs}. In \bibinfo{booktitle}{{\em WWW}}.
  \bibinfo{pages}{1215--1224}.
\newblock


\bibitem[\protect\citeauthoryear{Tang, Agarwal, O'Brien, and Meyer}{Tang
  et~al\mbox{.}}{2010}]%
        {Tang_kdd_2010}
\bibfield{author}{\bibinfo{person}{Diane Tang}, \bibinfo{person}{Ashish
  Agarwal}, \bibinfo{person}{Deirdre O'Brien}, {and} \bibinfo{person}{Mike
  Meyer}.} \bibinfo{year}{2010}\natexlab{}.
\newblock \showarticletitle{Overlapping Experiment Infrastructure: More,
  Better, Faster Experimentation}. In \bibinfo{booktitle}{{\em KDD}}.
  \bibinfo{address}{Washington, DC}, \bibinfo{pages}{17--26}.
\newblock


\bibitem[\protect\citeauthoryear{T{\"u}r}{T{\"u}r}{2007}]%
        {Tur_2007}
\bibfield{author}{\bibinfo{person}{G{\"o}khan T{\"u}r}.}
  \bibinfo{year}{2007}\natexlab{}.
\newblock \showarticletitle{Extending boosting for large scale spoken language
  understanding.}
\newblock \bibinfo{journal}{{\em Machine Learning (ML)\/}}
  \bibinfo{volume}{69}, \bibinfo{number}{1} (\bibinfo{year}{2007}),
  \bibinfo{pages}{55--74}.
\newblock


\bibitem[\protect\citeauthoryear{T{\"u}r, Wang, and Hakkani-T{\"u}r}{T{\"u}r
  et~al\mbox{.}}{2013}]%
        {Tur_signal_2013}
\bibfield{author}{\bibinfo{person}{G{\"o}khan T{\"u}r}, \bibinfo{person}{Ye-Yi
  Wang}, {and} \bibinfo{person}{Dilek~Z. Hakkani-T{\"u}r}.}
  \bibinfo{year}{2013}\natexlab{}.
\newblock \showarticletitle{TechWare: Spoken Language Understanding Resources
  [Best of the Web]}.
\newblock \bibinfo{journal}{{\em IEEE Signal Process. Mag. (SPM)\/}}
  \bibinfo{volume}{30}, \bibinfo{number}{3} (\bibinfo{year}{2013}),
  \bibinfo{pages}{187--189}.
\newblock


\bibitem[\protect\citeauthoryear{T{\"u}r, Wang, and Hakkani-T{\"u}r}{T{\"u}r
  et~al\mbox{.}}{2014}]%
        {Tur_2013}
\bibfield{author}{\bibinfo{person}{G{\"o}khan T{\"u}r}, \bibinfo{person}{Ye-Yi
  Wang}, {and} \bibinfo{person}{Dilek~Z. Hakkani-T{\"u}r}.}
  \bibinfo{year}{2014}\natexlab{}.
\newblock \showarticletitle{Understanding Spoken Language.}
\newblock \bibinfo{journal}{{\em Computing Handbook\/}} \bibinfo{volume}{3rd
  ed}, \bibinfo{number}{41} (\bibinfo{year}{2014}), \bibinfo{pages}{1--17}.
\newblock


\bibitem[\protect\citeauthoryear{Vinyals and Le}{Vinyals and Le}{2015}]%
        {VinyalsL15}
\bibfield{author}{\bibinfo{person}{Oriol Vinyals} {and}
  \bibinfo{person}{Quoc~V. Le}.} \bibinfo{year}{2015}\natexlab{}.
\newblock \showarticletitle{A Neural Conversational Model}.
\newblock \bibinfo{journal}{{\em CoRR\/}}  \bibinfo{volume}{abs/1506.05869}
  (\bibinfo{year}{2015}).
\newblock
\showURL{%
\url{http://arxiv.org/abs/1506.05869}}


\bibitem[\protect\citeauthoryear{Vinyals, Toshev, Bengio, and Erhan}{Vinyals
  et~al\mbox{.}}{2015}]%
        {vinyals2015show}
\bibfield{author}{\bibinfo{person}{Oriol Vinyals}, \bibinfo{person}{Alexander
  Toshev}, \bibinfo{person}{Samy Bengio}, {and} \bibinfo{person}{Dumitru
  Erhan}.} \bibinfo{year}{2015}\natexlab{}.
\newblock \showarticletitle{Show and tell: A neural image caption generator}.
  In \bibinfo{booktitle}{{\em CVPR}}. \bibinfo{pages}{3156--3164}.
\newblock


\bibitem[\protect\citeauthoryear{Weigel, Lu, Bailly, Oulasvirta, Majidi, and
  Steimle}{Weigel et~al\mbox{.}}{2015}]%
        {WeigelCHI2015}
\bibfield{author}{\bibinfo{person}{Martin Weigel}, \bibinfo{person}{Tong Lu},
  \bibinfo{person}{Gilles Bailly}, \bibinfo{person}{Antti Oulasvirta},
  \bibinfo{person}{Carmel Majidi}, {and} \bibinfo{person}{Jürgen Steimle}.}
  \bibinfo{year}{2015}\natexlab{}.
\newblock \showarticletitle{Iskin: flexible, stretchable and visually
  customizable on-body touch sensors for mobile computing}. In
  \bibinfo{booktitle}{{\em SIGCHI}}. \bibinfo{pages}{2991--3000}.
\newblock


\bibitem[\protect\citeauthoryear{Williams, Kiseleva, Crook, Zitouni, Awadallah,
  and Khabsa}{Williams et~al\mbox{.}}{2016a}]%
        {Williams_sigir_2016}
\bibfield{author}{\bibinfo{person}{Kyle Williams}, \bibinfo{person}{Julia
  Kiseleva}, \bibinfo{person}{Aidan Crook}, \bibinfo{person}{Imed Zitouni},
  \bibinfo{person}{Ahmed~Hassan Awadallah}, {and} \bibinfo{person}{Madian
  Khabsa}.} \bibinfo{year}{2016}\natexlab{a}.
\newblock \showarticletitle{Is This Your Final Answer? Evaluating the Effect of
  Answers on Good Abandonment in Mobile Search}. In \bibinfo{booktitle}{{\em
  SIGIR}}.
\newblock


\bibitem[\protect\citeauthoryear{Williams, Kiseleva, Crook, Zitouni, Awadallah,
  and Khabsa}{Williams et~al\mbox{.}}{2016b}]%
        {Williams_www_2016}
\bibfield{author}{\bibinfo{person}{Kyle Williams}, \bibinfo{person}{Julia
  Kiseleva}, \bibinfo{person}{Aidan~{C.} Crook}, \bibinfo{person}{Imed
  Zitouni}, \bibinfo{person}{Ahmed~Hassan Awadallah}, {and}
  \bibinfo{person}{Madian Khabsa}.} \bibinfo{year}{2016}\natexlab{b}.
\newblock \showarticletitle{Detecting Good Abandonment in Mobile Search}. In
  \bibinfo{booktitle}{{\em WWW}}. \bibinfo{pages}{495 -- 505}.
\newblock


\bibitem[\protect\citeauthoryear{Wu, Chin, Zhou, Wang, Guo, and Li}{Wu
  et~al\mbox{.}}{2012}]%
        {wu_2012}
\bibfield{author}{\bibinfo{person}{Liang Wu}, \bibinfo{person}{Alvin Chin},
  \bibinfo{person}{Yuanchun Zhou}, \bibinfo{person}{Xia Wang},
  \bibinfo{person}{Yonggang Guo}, {and} \bibinfo{person}{Jianhui Li}.}
  \bibinfo{year}{2012}\natexlab{}.
\newblock \showarticletitle{Context-aware prediction of user's first click}. In
  \bibinfo{booktitle}{{\em ContextDD}}.
\newblock


\bibitem[\protect\citeauthoryear{Xiang, Jiang, Pei, Sun, Chen, and Li}{Xiang
  et~al\mbox{.}}{2010a}]%
        {biaoxiang2010}
\bibfield{author}{\bibinfo{person}{Biao Xiang}, \bibinfo{person}{Daxin Jiang},
  \bibinfo{person}{Jian Pei}, \bibinfo{person}{Xiaohui Sun},
  \bibinfo{person}{Enhong Chen}, {and} \bibinfo{person}{Hang Li}.}
  \bibinfo{year}{2010}\natexlab{a}.
\newblock \showarticletitle{Context-Aware Ranking in Web Search}. In
  \bibinfo{booktitle}{{\em SIGIR}}.
\newblock


\bibitem[\protect\citeauthoryear{Xiang, Yuan, Zhao, Chen, Zhang, Yang, and
  Sun}{Xiang et~al\mbox{.}}{2010b}]%
        {Xiang_kdd_2010}
\bibfield{author}{\bibinfo{person}{Liang Xiang}, \bibinfo{person}{Quan Yuan},
  \bibinfo{person}{Shiwan Zhao}, \bibinfo{person}{Li Chen},
  \bibinfo{person}{Xiatian Zhang}, \bibinfo{person}{Qing Yang}, {and}
  \bibinfo{person}{Jimeng Sun}.} \bibinfo{year}{2010}\natexlab{b}.
\newblock \showarticletitle{Temporal Recommendation on Graphs via Long- and
  Short-term Preference Fusion}. In \bibinfo{booktitle}{{\em KDD}}.
  \bibinfo{pages}{723--732}.
\newblock


\bibitem[\protect\citeauthoryear{Yilmaz, Shokouhi, Craswell, and
  Robertson}{Yilmaz et~al\mbox{.}}{2010}]%
        {Yilmaz_cikm_2010}
\bibfield{author}{\bibinfo{person}{Emine Yilmaz}, \bibinfo{person}{Milad
  Shokouhi}, \bibinfo{person}{Nick Craswell}, {and} \bibinfo{person}{Stephen
  Robertson}.} \bibinfo{year}{2010}\natexlab{}.
\newblock \showarticletitle{Expected browsing utility for web search
  evaluation}. In \bibinfo{booktitle}{{\em CIKM}}. \bibinfo{pages}{1561--1564}.
\newblock


\bibitem[\protect\citeauthoryear{Young, Parsons, LeBeau, Tabak, Sewart, Stein,
  Kringelbach, and Craske}{Young et~al\mbox{.}}{2016}]%
        {Young_2016}
\bibfield{author}{\bibinfo{person}{Katherine~S Young},
  \bibinfo{person}{Christine~E Parsons}, \bibinfo{person}{Richard~T LeBeau},
  \bibinfo{person}{Benjamin~A Tabak}, \bibinfo{person}{Amy~R Sewart},
  \bibinfo{person}{Alan Stein}, \bibinfo{person}{Morten~L Kringelbach}, {and}
  \bibinfo{person}{Michelle~G Craske}.} \bibinfo{year}{2016}\natexlab{}.
\newblock \showarticletitle{Sensing Emotion in Voices: Negativity Bias and
  Gender Differences in a Validation Study of the Oxford Vocal (‘OxVoc’)
  Sounds Database}.
\newblock \bibinfo{journal}{{\em American Psychological Association\/}}
  (\bibinfo{year}{2016}).
\newblock


\bibitem[\protect\citeauthoryear{Zhu, Chen, Xiong, Yu, Cao, and Tian}{Zhu
  et~al\mbox{.}}{2015}]%
        {zhu2015mining}
\bibfield{author}{\bibinfo{person}{Hengshu Zhu}, \bibinfo{person}{Enhong Chen},
  \bibinfo{person}{Hui Xiong}, \bibinfo{person}{Kuifei Yu},
  \bibinfo{person}{Huanhuan Cao}, {and} \bibinfo{person}{Jilei Tian}.}
  \bibinfo{year}{2015}\natexlab{}.
\newblock \showarticletitle{Mining mobile user preferences for personalized
  context-aware recommendation}.
\newblock \bibinfo{journal}{{\em TIST\/}} \bibinfo{volume}{5},
  \bibinfo{number}{4} (\bibinfo{year}{2015}), \bibinfo{pages}{58}.
\newblock


\end{thebibliography}
